\documentclass[aip, cha, reprint]{revtex4-1}

\usepackage{amsmath,amsfonts,amssymb,amsthm,graphicx,bbm,enumerate,times,color}
\usepackage{mathtools}
\usepackage{wasysym}

\newcommand{\e}{{\rm e}}

\newcommand{\pik}{Potsdam Institute for Climate Impact Research, 14473 Potsdam, Germany}
\newcommand{\hu}{Department of Physics, Humboldt University, 12489 Berlin, Germany}

\begin{document}
\title{Phase space reconstruction for non-uniformly sampled noisy time series}

\author{Jaqueline Lekscha}
\affiliation{\pik}
\affiliation{\hu}
\author{Reik V. Donner}
\affiliation{\pik}

\date{\today}

\begin{abstract}
Analyzing data from paleoclimate archives such as tree rings or lake sediments offers the 
opportunity of inferring information on past climate variability. Often, such data sets are 
univariate and a proper reconstruction of the system's higher-dimensional 
phase space can be crucial for further analyses. 
In this study, we systematically compare the methods of time delay embedding and differential embedding
for phase space reconstruction.
Differential embedding relates the system's higher-dimensional 
coordinates to the derivatives of the measured time series. For implementation, this requires 
robust and efficient algorithms to estimate derivatives from noisy and possibly non-uniformly sampled 
data. For this purpose, we consider several approaches: (i) central differences adapted to irregular 
sampling, (ii) a generalized version of discrete Legendre coordinates and (iii) the concept of 
Moving Taylor Bayesian Regression.
We evaluate the performance of differential and time delay embedding by studying two paradigmatic model 
systems -- the Lorenz and the R\"ossler system. More precisely, we compare geometric properties
of the reconstructed attractors to those of the original attractors by applying recurrence network analysis.
Finally, we demonstrate the potential and the limitations of using the different phase 
space reconstruction methods in combination with windowed recurrence network analysis for inferring
information about past climate variability. This is done by analyzing two well-studied paleoclimate 
data sets from Ecuador and Mexico. We find that studying the robustness of the results when varying the 
analysis parameters is an unavoidable step in order to make well-grounded statements on climate 
variability and to judge whether a data set is suitable for this kind of analysis. 
\end{abstract}

\maketitle

\begin{quotation}
Nonlinear methods such as recurrence quantification analysis and recurrence network analysis offer the 
prospect of gaining valuable insights into past climate variability. 
In order to apply these methods, the phase space of the underlying dynamical system has to be 
reconstructed if the system dynamics are expected to be higher-dimensional than the measured data. 
In the paleoclimate context, observations are usually univariate while the dynamics of the climate 
system cannot be assumed to be low-dimensional. 
Additionally, time series of paleoclimate proxies are often non-uniformly sampled and subject to noise, 
giving rise to challenges for phase space reconstruction and raising the question how reliable
results are when applying nonlinear methods.
By systematically studying different phase space reconstruction techniques and combining them with windowed 
recurrence network analysis, we want to address this problem.
We find that not all data sets are suited for recurrence network analysis and that the robustness of the
results has to be studied before drawing conclusions about past climate variability.  
\end{quotation}

\section{Introduction}
Reconstructing a system's phase space from measured data is a standard procedure in nonlinear time series 
analysis. In particular, finding a proper embedding is a key step in many applications. 
Among others, recurrence-based methods often make use of embedded time series and have been proven useful in studying dynamical transitions in a variety of disciplines, 
ranging from medical to paleoclimate data~\cite{Marwan2007, Donges2011}. 
For example,
when aiming at detecting and characterizing nonlinear regime shifts in past climate variability
by analyzing data from paleoclimate archives such as tree rings, lake sediments or speleothems, the 
reconstruction of the system's higher-dimensional phase space is unavoidable. 
This is, because the dynamics of the climate system are not expected to be low-dimensional but the
measured time series are commonly univariate. Also, proxy variables can be assumed to have been 
influenced by more than one climate variable.
Since such data are 
inherently subject to noise and very often non-uniformly sampled, methodological difficulties arise.
For example, time delay embedding is not well defined for non-uniformly sampled data. Moreover, common 
approaches to estimate appropriate embedding parameters rely on regularly sampled time 
series~\cite{Rehfeld2011}.
Still, such kind of analyses constitute an important step towards better
understanding the climate of the past. This is not only a historically relevant task but can also serve
to assess major implications for modern civilizations in the context of recent climate change. 
It has been shown that climate variability and extremes can trigger migration~\cite{Donges2015a}, 
armed conflict~\cite{Schleussner2016} and the development of civilization~\cite{Donges2015a}, 
the most prominent example being the rise and fall of the Classic Maya Culture~\cite{Kennett2012,Carleton2017}. 

To gain additional insights into the underlying dynamical principles of the climate system, a reliable 
analysis framework is required. We here aim at contributing to the further development of such a framework 
by systematically studying different methods to reconstruct the phase space of a system from a measured 
univariate time series that may be non-uniformly sampled and subject to noise. In particular, we 
study how those different phase space reconstructions behave when combining them with
windowed recurrence network analysis. 
To do so, we compare the method of time delay embedding with that of differential embedding and present 
several approaches of estimating derivatives from noisy, non-uniformly sampled data in 
Sec.~\ref{sec:methods}. We then reconstruct the phase space of two paradigmatic model systems 
and compare their geometric properties to those of the original attractors using recurrence network 
analysis in Sec.~\ref{sec:ms}.
Finally, we apply the different phase space reconstruction methods in combination with windowed recurrence 
network analysis to two well-studied real-world paleoclimate data sets and test the robustness of
this analysis framework when varying the different analysis parameters in Sec.~\ref{sec:pcts}.
We summarize our insights and conclusions in Sec.~\ref{sec:conclusion}.

\section{Methods} \label{sec:methods}

\subsection{Phase space reconstruction}
Attractors of dynamical systems can be reconstructed topologically equivalent from data as has been 
shown in the embedding theorems by Whitney~\cite{Whitney1936}, Takens~\cite{Takens1980} and 
Ma\~{n}\'{e}~\cite{Mane1981}.
Takens and Packard
independently suggested differential embedding and time delay 
embedding
as two methods to reconstruct the higher-dimensional phase space of a system from a measured,
univariate time series~\cite{Takens1980,Packard1980}. 

\emph{Time delay embedding} substitutes the non-observed essential coordinates of the
system by delayed versions of the measured time series, i.\,e., if the time series $x(t_i)$ was 
measured at times $t_i, i=1,\dots, N$, the reconstructed state vectors are given by
\begin{align}
x(t_i) \rightarrow \left\lbrace x(t_i), x(t_i-\tau), \dots, x(t_i-(m-1)\tau)\right\rbrace \label{eq:tde}
\end{align}
with $\tau$ some suitable time delay and $m$ the embedding dimension. 
In a similar way, \emph{differential embedding} relates
the higher-dimensional coordinates to the respective derivatives up to order $m-1$,
\begin{align}
x(t_i) \rightarrow \left\lbrace x(t_i), \dfrac{d x(t_i)}{d t}, \dots, \dfrac{d^{m-1}x(t_i)}{d t^{m-1}}\right\rbrace. \label{eq:diffemb}
\end{align}

Takens proved that for regularly sampled and noise free data, the attractors reconstructed using
time delay or differential embedding are diffeomorphic to the original attractor if the 
embedding dimension is larger than twice the dimension of the original attractor~\cite{Takens1980}. 
The work of Packard et al.~\cite{Packard1980} was more practically inspired, searching for independent 
coordinate representations from experimental data and suggesting both the method of delays and the 
method of derivatives. 

Since then,  much work has 
focused on how to choose the delay time and embedding dimension for given data.
Practically, the dimension of the system's attractor is not known and any measurement
process will be subject to noise. For noise-free data, any delay time $\tau$ can be chosen, whereas
in the presence of noise the choice of the embedding delay can be crucial~\citep{Casdagli1991}. 
The first zero of the autocorrelation 
or the first minimum of the mutual information have been suggested as practically applicable 
criteria~\cite{Kantz2004,Abarbanel1993,Fraser1986}. 
To deal with noise, principal component embedding has been 
developed~\cite{Gibson1992,Casdagli1991}. For the 
embedding dimension, the false nearest neighbor algorithm has
been shown to provide a reasonable estimate~\cite{Kennel1992}. 
A recent approach avoids the choice of dimension by 
using an infinite dimensional embedding with weighted coordinates, introducing a scaling 
factor~\cite{Hirata2017}. 

The above discussion shows that the choice of embedding parameters is a subject of ongoing research.   
In this context, the problem of non-uniform sampling of the data has not yet been addressed 
sufficiently. In this case, time 
delay embedding suffers from conceptual problems. Not only standard estimators for the 
autocorrelation and mutual information rely on regularly sampled time 
series~\cite{Ozken2015,Rehfeld2011,Rehfeld2014}, also the embedding 
itself is not well defined for irregular sampling such that interpolation of the data becomes 
unavoidable. In turn, differential embedding does not suffer from those conceptual problems and the choice 
of derivatives as higher-dimensional coordinates may seem more natural than that of delays, in particular,
when thinking of the dynamics of a system to be generated by differential equations.
In addition, it obeys only one characteristic parameter ($m$) instead of two ($m$, $\tau$), which may be 
beneficial in real-world cases. However, differential embedding
has not yet been used often in practice as the robust numerical estimation of derivatives from noisy 
data is challenging.

\subsection{Estimating derivatives}
In the following, we discuss different techniques to estimate derivatives from noisy and possibly 
non-uniformly sampled data. That is, given a time series $x_i = x(t_i)$ for $i=1,\dots, N$ and 
non-uniform sampling intervals $\Delta t_{i} = t_{i+1}-t_i$, we want to approximate the time 
derivatives 
\begin{align}
\frac{d^j x}{d t^j}\bigg\vert_{t=t_i} &\equiv \frac{d^j}{d t^j} x_i 
\end{align}
at all times $t_i$ and for orders $j$ up to some order $j_{\max}$.

\subsubsection{Central differences}
The first and probably simplest approach to numerically estimate derivatives is to approximate the 
derivative by differences.
For regular sampling the central difference quotient is given as
\begin{align}
\frac{d}{d t} x_i \approx \frac{\Delta x_i}{\Delta t} &= \dfrac{x_{i+1}-x_{i-1}}{2\Delta t}. \label{eq:cdiffreg}
\end{align}
Taking non-uniform sampling into account, this formula can be generalized to
\begin{align}
\frac{\Delta x_i}{\Delta t} = \dfrac{\Delta t_{i-1}^2 x_{i+1} - (\Delta t_{i-1}^2 - \Delta t_{i}^2) x_i - \Delta t_{i}^2 x_{i-1}}{\Delta t_{i-1} \Delta t_{i} ( \Delta t_{i-1} + \Delta t_{i})},\label{eq:cdiff}
\end{align}
which reduces to~\eqref{eq:cdiffreg} when $\Delta t_{i} = \Delta t_{i-1}$. 
To derive the above expression, we approximate 
the values of the neighboring points of our target point $x_i$ using a Taylor expansion up to second order:
\begin{align}
x_{i-1} &\approx x_i - \Delta t_{i-1} \frac{d}{d t} x_i + \dfrac{\Delta t_{i-1}^2}{2}\frac{d^2}{d t^2} x_i, \nonumber \\
x_{i+1} &\approx x_i + \Delta t_{i} \frac{d}{d t} x_i + \dfrac{\Delta t_{i}^2}{2}\frac{d^2}{d t^2} x_i . \nonumber
\end{align}
Eliminating the term with the second derivatives from this system of equations and solving for 
$\frac{d}{d t} x_i$ gives the desired result~\eqref{eq:cdiff}. Higher order derivatives can be obtained by 
repeatedly applying Eq.~\eqref{eq:cdiff}.  

\subsubsection{Discrete Legendre polynomials}
A refined approach to estimate derivatives suggested by Gibson et al.~\cite{Gibson1992} uses discrete Legendre polynomials. This concept can be generalized to 
irregular sampling and relates the $j$th derivative at $x_i$ to a weighted sum of the
$p$ nearest points to each side of $x_i$ as
\begin{align}
\dfrac{d^j}{d t^j} x_i &\approx \dfrac{j !}{c_{j,p}(\Delta t_{i,n})} \sum_{n=-p}^p r_{j,p,n}^{(i)} x_{i+n} \label{eq:dLp}
\end{align}
with $\Delta t_{i,n} = t_{i+n} - t_i$ and
\begin{align}
c_{j,p}(\Delta t_{i,n}) &= \sum_{n=-p}^p (\Delta t_{i,n})^j r_{j,p,n}^{(i)}. \nonumber
\end{align}
The weights are given by the discrete Legendre polynomials $r_{j,p,n}^{(i)} = r_{j,p}(\Delta t_{i,n})$ that can be calculated recursively by
the relation
\begin{align}
r_{j,p,n}^{(i)} &= \dfrac{1}{c_j p^j}\left[\Delta t_{i,n}^j - \sum_{k=0}^{j-1} r_{k,p,n}^{(i)} \sum_{l=-p}^p \Delta t_{i,l}^j r_{k,p,l}^{(i)}\right] \label{eq:rec-relation}
\end{align}
for $2p\geq j$ with $r_{0,p,n}^{(i)} = 1/c_0$. 
The normalization constants $c_j$ can be determined by the conditions
\begin{align}
\sum_{n=-p}^p (r_{j,p,n}^{(i)})^2 &= 1. \label{eq:normalization}
\end{align}
The derivation of~\eqref{eq:dLp} can be found in the appendix~\ref{app-dLp}.
It should be noted that the discrete Legendre polynomials are not a discretization of the common 
Legendre polynomials. Instead, for $p\rightarrow\infty$, they converge to the latter.
By changing the parameter $p$ (the number of neighbors to each side that are included for 
estimating the derivatives) it is possible to control the smoothing of the data. That is, some
noise can be averaged out which makes this procedure more robust with respect to noise than other 
methods. For estimating the optimal value of $p$, a procedure has been suggested by
Gibson et al.\cite{Gibson1992}. In general, 
we recommend choosing a $p$ of the order of the embedding dimension, i.\,e. of the order of
the highest derivative that needs to be estimated. In the limit case of 
choosing $p$ as small as possible, this approach reduces to a central difference quotient.

\subsubsection{Moving Taylor Bayesian Regression}
The concept of Moving Taylor Bayesian Regression (MoTaBaR) introduced by 
Heitzig~\cite{Heitzig2013} can 
be used to estimate the value of some function $f$ and its derivatives of order smaller than $p$ at some specific 
position of interest $\xi$. To do so, a set of $N$ measured data points, prior beliefs about the 
variability of argument and measurement error and about the variability of $f$ and some of its derivatives 
are required. In the context of paleoclimate data, the argument and measurement errors correspond to
the time and measurement uncertainties.
Using the measured data and Bayesian updating, posterior distributions can be obtained from the prior 
distributions and the posterior mean and variance can be used as estimators of the value and uncertainty
of the function and its derivatives at $\xi$. To be more precise, we here assume that the measured data set 
$\left\lbrace x_i,y_i\right\rbrace_{i=1}^N$ is one-dimensional, 
that the argument and measurement errors ($\gamma_i$ and $\epsilon_i$) are Gaussian and that we do not have
prior information about the variability of the function $f$ and its derivatives. 
A local Taylor approximation at the position of interest $\xi$ relates the measured values to the 
derivatives  $\varphi^{\alpha} = \frac{d^{\alpha} f(\xi)}{d x^{\alpha}}$ as
\begin{align}
y_i &= \sum_{\alpha=0}^{p-1} X_{i,\alpha} \varphi^{\alpha} + r_i + \epsilon_i.\nonumber
\end{align}
The $X_{i,\alpha}$ are given by
\begin{align}
X_{i,\alpha} &= \dfrac{(x_i - \gamma_i - \xi)^{\alpha}}{\alpha!} \label{eq:mtb-X}
\end{align}
and $r_i$ is the Taylor residual
\begin{align}
r_i &= X_{i,p} \psi_{i},\nonumber\\ 
\psi_{i} &= \dfrac{d^p}{d x^p} f(\xi + \lambda_{i} (x_i-\gamma_i-\xi))\nonumber
\end{align}
with $\lambda_{i}\in [0,1]$. The procedure to estimate the posterior mean $\tilde{\mu}_{\varphi}$ and 
variance $\tilde{\Sigma}_{\varphi}$ of $f$ and its derivatives for non-informative priors, i.\,e.\,\,prior
variance $\Sigma_{\varphi}^{-1} = 0$ and prior mean of $\psi$, $\mu_{\psi} = 0$, is as follows: 
\begin{itemize}
\item[(i)] Calculate $X_{i,\alpha}$ for all $i$ and $\alpha<p$ from $x$, $\gamma$ and $\xi$ 
using~\eqref{eq:mtb-X}. 
\item[(ii)] Calculate $\left(\Sigma_{r}\right)_{i,j}$ from the prior belief about $\left(\Sigma_{\psi}\right)_{i,j}$
\begin{align}
\left(\Sigma_{r}\right)_{i,j} &= X_{i,p} \left(\Sigma_{\psi}\right)_{i,j} X_{j,p}.\nonumber
\end{align}
\item[(iii)] Calculate $W = (\Sigma_r + \Sigma_{\epsilon})^{-1}$. 
\item[(iv)] Obtain the posterior variance and mean as
\begin{align}
\tilde{\Sigma}_{\varphi} &= (X'WX)\nonumber\\
\tilde{\mu}_{\varphi} &= \tilde{\Sigma}_{\varphi} (X'Wy).\nonumber
\end{align}
\end{itemize}
If the distribution of the argument errors $\rho(\gamma|x)$ is known, the final estimators $\hat{\varphi}$
and $\hat{\Sigma}_{\varphi}$ can be obtained by integrating over all possible argument errors.
A detailed and more general description of the method can be found in~\cite{Heitzig2013}.

\subsection{Recurrence network analysis} \label{subsec:rna}
Recurrences in phase space have been successfully used for characterizing system 
dynamics in many fields of application~\cite{Marwan2007}. The underlying concept of recurrence plots
relies on the recurrence matrix $R_{i,j}$
\begin{align}
R_{i,j}(\epsilon) &= \theta\left(\epsilon - \| \vec{x}_i - \vec{x}_j \|\right)\label{eq:rm}
\end{align}
that quantifies at which points $t_j$ in time the observed trajectory of the system is closer
to some previous state $\vec{x}_i = \vec{x}(t_i)$ than a given threshold distance
$\epsilon$ with respect to some metric $\|\cdot\|$ in phase space. Here, $\theta(\cdot)$ is the Heaviside 
function and $\vec{x}_i$ is the state vector of the system at time $t_i$.
To quantify distances, we use the supremum norm
\begin{align}
\| \vec{z} \|_{\infty} &= \max_{j=1,\dots, m}\left\lbrace z^{(j)}\right\rbrace\nonumber
\end{align}
with $m$ the dimension of the time series.

Beyond the characterization of recurrences in phase space, time series analysis using complex network techniques has received considerable attention in the 
past few years. Different classes of networks have been suggested to characterize the dynamics of a 
system from a time series~\cite{Donner2010a,Donner2011}. Among these different concepts,
recurrence networks integrate the two aspects of
recurrences and time series analysis using complex networks by reinterpreting
the recurrence matrix as the adjacency matrix of a complex network as 
\begin{align}
A_{i,j}(\epsilon) &= R_{i,j}(\epsilon) - \delta_{i,j},\label{eq:am}
\end{align}
where the delta function $\delta_{i,j}$ excludes self-connections. That is, the nodes of the recurrence 
network correspond to the state vectors of the system under study that can be identified with a certain 
time of observation. A link between 
two nodes is established when the states are closer than the threshold $\epsilon$ in phase space.

It should be noted that recurrence network analysis solely captures geometric properties of the system in 
phase space and is invariant under re-ordering of the nodes. Specifically, the network structure depends on 
the geometric characteristics of the system's attractor and can be
characterized using measures like the average path length $\mathcal{L}$ and
network transitivity $\mathcal{T}$~\cite{Donner2010,Donner2011b}. This makes recurrence network analysis 
particularly suitable for comparing different attractor reconstructions in phase space.
The average path length is defined as
\begin{align}
\mathcal{L} &= \dfrac{2}{N(N-1)} \sum_{i<j} l_{i,j} \label{eq:avpathlength}
\end{align}
with $l_{i,j}$ denoting the length of the shortest path between node $i$ and node $j$. Low average 
path lengths have been attributed to more regular dynamics of the system while
high values indicate more chaotic dynamics~\cite{Donner2010}.
The network transitivity is defined as 
\begin{align}
\mathcal{T} &= \dfrac{\sum_{v,i,j} A_{v,i}A_{i,j}A_{j,v}}{\sum_{v,i,j} A_{v,i}A_{j,v}}, \label{eq:transitivity}
\end{align}
which can be interpreted as the probability that two neighbors $i$ and $j$ of a randomly chosen node $v$ 
in the network are mutually connected. High values of the transitivity correspond to lower-dimensional
dynamics while low values relate to higher-dimensional dynamics~\cite{Donner2010}. 
In this context, transitivity has been demonstrated to constitute a generalized notion of dimensionality
of a chaotic attractor~\cite{Donner2011b}.

\section{Model systems} \label{sec:ms}
To study the potential of the different ways to reconstruct the phase space of a system from noisy and 
non-uniformly sampled data, we first consider two model systems, the Lorenz~\cite{Lorenz1963} 
and the R\"ossler system~\cite{Rossler1976}, which are given as
\begin{align}
\dot{x} &= a(y-x)\nonumber\\
\dot{y} &= x(b-z)-y\nonumber\\
\dot{z} &= xy-cz\nonumber
\end{align}
and
\begin{align}
\dot{x} &= -y-z\nonumber\\
\dot{y} &= x + dy\nonumber\\
\dot{z} &= e + z(x-f),\nonumber
\end{align}
respectively.
To obtain a high-resolution reference solution, we numerically integrate the respective set of 
differential equations with a high sampling rate ($d t = 0.001$) with regular sampling. 
As parameters and initial conditions, we
chose  $a=10, b=28, c=8/3$, $x_0=-8.0, y_0 = 8.0, z_0 = 27.0$ for the Lorenz 
and $d=0.15$, $e=0.2$, $f=10.$, $x_0=0.5$, $y_0 = 0.0$, $z_0 = 0.0$ for the R\"ossler system.
These choices of parameters correspond to the canonical choices that lead to chaotic solutions.

\subsection{Analysis procedure}
The basic idea is to compare some characteristics of recurrence networks constructed from the 
reference attractors and using different phase space reconstructions.
To do so, we first construct non-uniformly sampled and noisy time series from the reference 
series. This is done in a two-step procedure. First, we randomly draw $N-1$ time intervals from a gamma distribution
\begin{align}
P_{\Gamma}(t) &= t^{k-1} \dfrac{\exp(-t/\theta)}{\theta^k \Gamma(k)} \label{eq:gamma-dist}
\end{align}
where $\theta$ is the scale and $k$ is the shape parameter of the distribution. By using the
corresponding values of the $x$-coordinate of the respective model system at the resulting times, 
we end up with a non-uniformly sampled time series
of length $N$. Specifically, we employ $500$ independent random permutations of the gamma distributed
sampling intervals to obtain $500$ different realizations
of the non-uniformly sampled time series. 

In a second step, we add white noise to the data 
\begin{align}
x_i &\rightarrow x_i + \eta_i\nonumber
\end{align}
where the noise $\eta$ is chosen according to a normal distribution
\begin{align}
p(x) &= \dfrac{1}{\sqrt{2\pi \sigma^2}} \e^{-\frac{(x-\mu)^2}{2\sigma^2}} \label{eq:normal-dist}
\end{align}
with mean $\mu$ and variance $\sigma^2$. The resulting time series exhibits some typical characteristics
of measured time series and is then used to reconstruct the systems' attractors in phase space using time 
delay embedding on the one hand and differential embedding with different methods to estimate the 
derivatives on the other hand. Note that the effect of noise on attractor reconstructions from regularly 
sampled time series has already been studied elsewhere~\cite{Xiang2012,Jacob2016}.

As time delay embedding requires regular sampling, we first interpolate the data back to
regular sampling using either linear or cubic spline interpolation. For differential embedding, we directly 
use the non-uniformly sampled data. Additionally, we consider the two interpolated data sets and 
also compare the situations with and without scaling each coordinate of the embedded data to unit 
variance. Note that for time delay embedding, the variance of all reconstructed coordinates is by 
definition of the same order while for differential embedding this is not necessarily the case. 
The scaling to unit variance accounts for that fact. 

From each
embedded time series, we finally construct recurrence networks and calculate network transitivity 
$\mathcal{T}$ and average path length $\mathcal{L}$ to characterize the attractors. 
We also calculate the corresponding network measures for regular reference solutions $(x,y,z)$ of equal 
length $N$ and denote them as $\mathcal{T}_{\text{ref}}$ and $\mathcal{L}_{\text{ref}}$, respectively.
This is done by subsampling the original high resolution time series according to the average sampling 
time of the lower resolution non-uniformly sampled time series.
To quantify the quality of the attractor reconstruction, we use the mean and standard deviation of the 
relative difference of the network measures between the reference and the reconstructed attractor 
\begin{align}
\Delta \mathcal{T} &= \dfrac{\lvert\mathcal{T}_{\text{ref}}-\mathcal{T}\rvert}{\mathcal{T}_{\text{ref}}}, \qquad \Delta \mathcal{L} = \dfrac{\lvert\mathcal{L}_{\text{ref}}-\mathcal{L}\rvert}{\mathcal{L}_{\text{ref}}}\nonumber
\end{align}
taken over all $500$ realizations of the non-uniform sampling.

\subsection{Results}

\begin{figure*}
\centering
\includegraphics[width=0.99\linewidth]{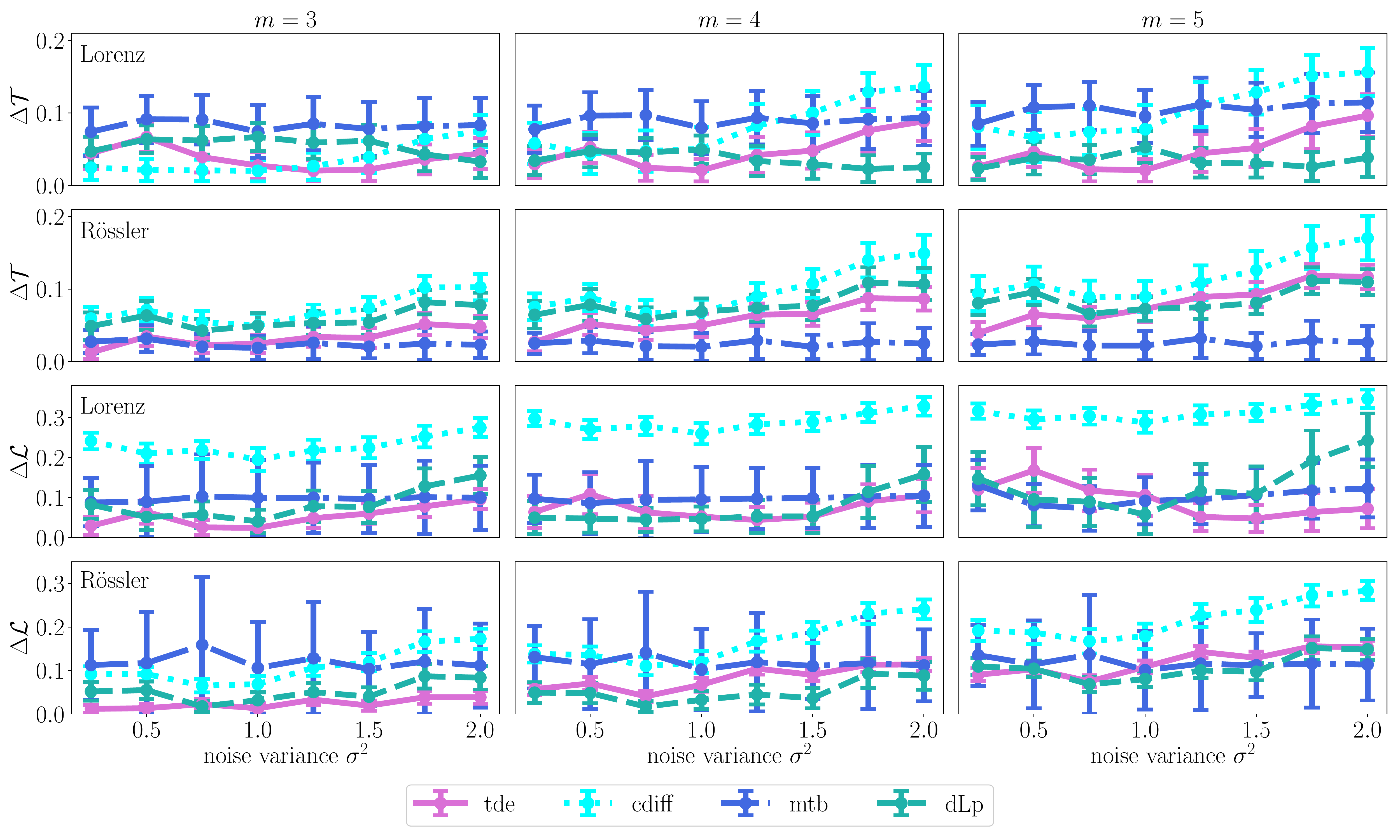}
\caption{Mean and standard deviation of $\Delta \mathcal{T}$ and $\Delta \mathcal{L}$ as a function of the noise variance $\sigma^2$ for the Lorenz and the R\"ossler system and increasing embedding dimension for different attractor reconstructions. Purple solid lines: time delay embedding (tde) for $\tau = 2 \left\langle\Delta t\right\rangle$ (Lorenz) and $\tau = \left\langle\Delta t\right\rangle$ (R\"ossler) and linear interpolation. Light blue dotted lines: central differences (cdiff) scaled to unit variance for cubic spline interpolation. Dark blue dash-dotted lines: MoTaBaR (mtb) based derivative estimates scaled to unit variance and internally interpolated. Cyan dashed lines: discrete Legendre polynomials (dLp) for $p=4$ ($m=3$) and $p=6$ ($m=4,5$) for cubic spline interpolation scaled to unit variance (Lorenz) and without scaling (R\"ossler).}
\label{fig:noise}
\end{figure*} 

\begin{figure*}
\centering
\includegraphics[width=0.99\linewidth]{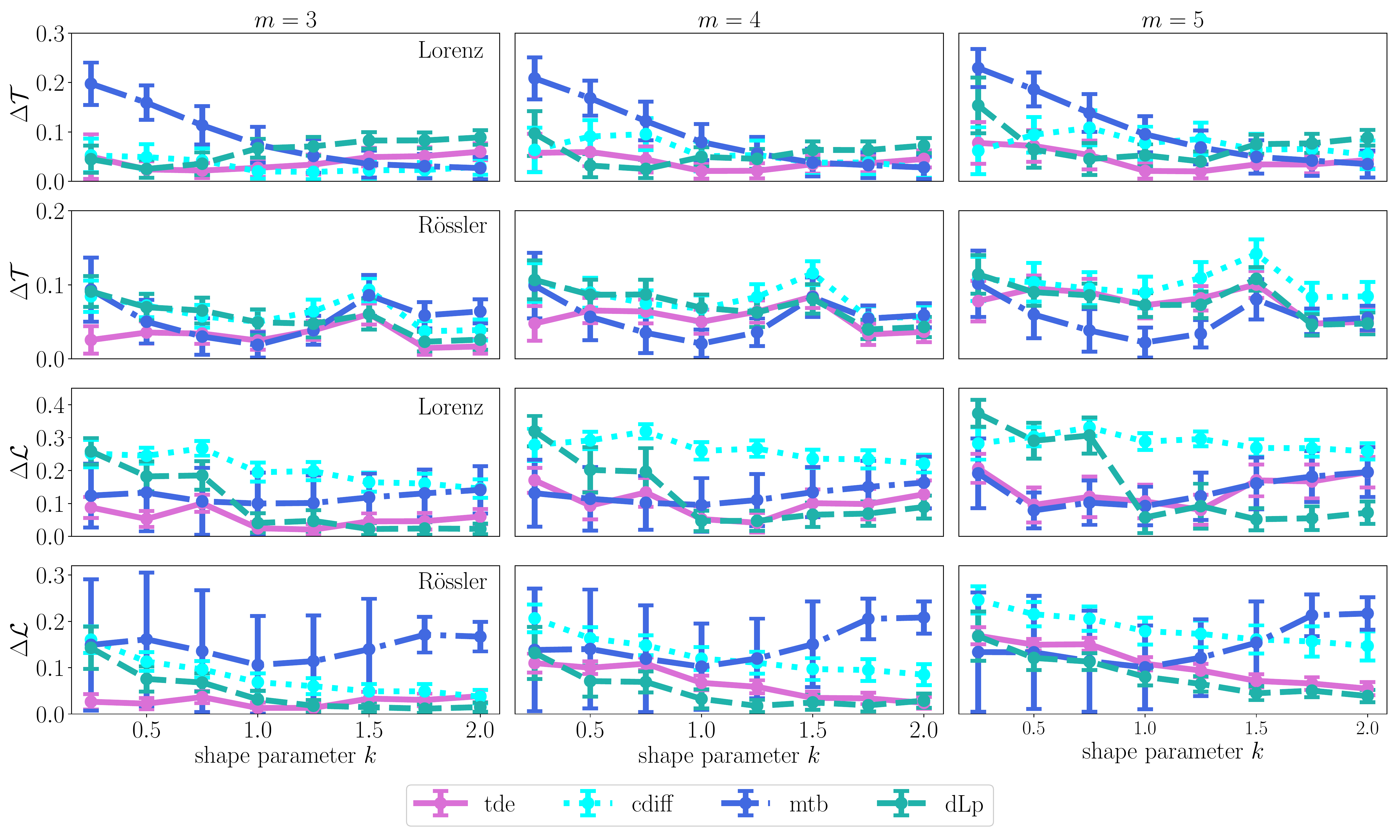}
\caption{Same as in Fig.~\ref{fig:noise} for the dependence on the shape parameter $k$ of the sampling interval distribution.}
\label{fig:shape}
\end{figure*} 
We now study how well the different attractor reconstruction techniques perform 
when varying (i) the noise level and (ii) the shape parameter of the gamma distribution
that determines the non-uniform sampling. For this, we sample ensembles of $N=500$ points each from the 
reference solutions as described in the previous section. The sampling intervals are distributed 
according to the gamma 
distribution~\eqref{eq:gamma-dist} with scale parameter $\theta = 8.0$ and varying shape parameter 
$k \in [0.25,2.0]$. The additive noise is given by Eq.~\eqref{eq:normal-dist} with mean $\mu = 0$
and variance $\sigma^2$ varying in the interval $[0.25,2.0]$. 

For time delay embedding, we chose the linearly interpolated data and a delay 
time of $\tau= \left\langle\Delta t\right\rangle$ for the R\"ossler and of $\tau=2 \left\langle\Delta t\right\rangle$ for the Lorenz system with $\left\langle\Delta t\right\rangle$ being
the corresponding average sampling time. This choice of delay times is motivated by 
calculating the autocorrelation of the time series as a function of the delay and choosing the first zero
as delay time $\tau$~\cite{Kantz2004}. Also, we found that these choices of the delay time correspond to 
those that give the best results for the attractor reconstruction. 
We chose linear interpolation as this is the most commonly used technique when reconstructing the phase
space of a system from non-uniformly sampled data using time delay embedding. 

For differential embedding using central differences, a procedure first interpolating the data to regular 
sampling via cubic splines 
and then rescaling the coordinates to unit variance was found to perform best. 
For MoTaBaR, we focus on the derivative estimates that were 
interpolated to regular sampling using the internal MoTaBaR interpolation routine and including 
$N_{mtb} = 20$ data points to estimate the derivatives.
Here, the data were again scaled to unit variance. Finally, for the discrete Legendre polynomials, 
we also found the best results for cubic spline interpolation. 
For the Lorenz system, additionally rescaling the 
data to unit variance greatly improved the results while for the R\"ossler system, the results were best 
without rescaling. A possible reason for this is that the R\"ossler system exhibits one outstanding 
coordinate ($z$) with values of a different order of magnitude than the other coordinates. As the
number of neighboring points that are included to each side when estimating the derivatives, we chose
$p=4$ for embedding dimension $m=3$ and $p=6$ for $m=4$ and $m=5$. As already 
mentioned, the choice of $p$ can be justified using a recursive algorithm proposed in~\cite{Gibson1992}. 
In our case this suggests $p$ of the order $3$. As this estimation
procedure is independent of the embedding dimension and we require estimates of higher order derivatives,
we prefer to choose $p$ accordingly a bit higher. Another approach to estimate a reasonable
value for $p$ for oscillating data is to choose $p$ such, that the points taken into account to both sides 
together cover approximately a quarter of an oscillation~\cite{Gibson1992}. For very noisy data, applying this approach can
however be 
difficult in practice as identifying the system's oscillations from noisy data can be challenging. 
For the model
systems considered here, the latter method would result in a choice of $p$ around $6$. 
Our choice of $p$ is thus in accordance with the above considerations. Also, we find that those choices
perform quite well compared to other 
values of $p$. In particular, we find that for lower/higher embedding dimensions a slightly lower/higher 
choice of $p$ performs better (not shown). 

When increasing the noise level (Fig.~\ref{fig:noise}), we expect 
the results to gradually get worse. Also, for higher embedding dimensions and differential embedding, the
noise effect should amplify as higher-order derivatives need to be estimated which are more sensitive to
noise. For time delay embedding, this trend is visible in most cases. Still, this approach performs 
quite well for high noise levels, in particular for the difference of the average path length in the 
case of $m=3$ and in the case of $m=6$ for the Lorenz system. Also for differential embedding using 
central differences, the reconstructions tend to differ more from the reference attractor
for higher noise level. However, this method generally performs worse than the other methods. 
For differential embedding using MoTaBaR, we observe a relatively constant difference to the 
reference attractors' properties but a higher standard deviation among the $500$
different realizations. This method performs particularly well compared to the other approaches for the 
difference in transitivity and the R\"ossler system. Also, MoTaBaR in conjunction with 
irregular sampling and 
using the internal interpolation shows in all cases a very similar behavior, though the interpolated 
case always gives slightly better results (not shown). When using differential embedding with discrete Legendre 
polynomials, we sometimes observe the expected increase of the difference to the reference case when
increasing the noise level. However, in other cases, the results hardly depend on the noise level, 
in particular for the difference in 
transitivity for the Lorenz system and also up to a high noise level for the difference in average path 
length for the R\"ossler system.

When changing the shape parameter $k$ of the gamma distribution (Fig.~\ref{fig:shape}), 
we expect better performance for higher values of the parameter as in this
case, the distribution is more centered and thus closer to regular sampling. For time delay embedding, 
this expected trend is visible in some cases but not in all. 
Still, it performs relatively good for most cases.
Differential embedding using central differences for derivative estimation provides
rather constant results when varying the shape parameter of the gamma distribution. But, except for the
difference to the transitivity of the Lorenz system where it exhibits reasonable results for small values of the 
shape parameter, this approach does not perform very well compared to the other methods. 
Differential embedding using MoTaBaR derivative estimates shows the expected 
trend very clearly for the difference to the reference transitivity. For the difference in the average 
path length, it first improves performance with increasing the shape parameter, then, when further 
increasing it, the performance gets worse again. For differential embedding using discrete Legendre 
polynomials to estimate the derivatives, we mostly find the expected trend and particularly notice a
large difference in performance for the average path length in the Lorenz 
system between $k=0.75$ and $k=1.0$.

Overall, we have found that time delay embedding 
using linear interpolation performs quite well but often differential embedding using 
discrete Legendre polynomials or MoTaBaR performs even better. 
In contrast to this, differential embedding with estimating the derivatives by
central differences does not perform very well in almost all cases. Hence, we conclude that more
sophisticated methods should be applied to estimate derivatives from noisy and non-uniformly sampled
data. Also, we have found that in general, the recurrence network transitivity is slightly closer to the 
reference transitivity than the average path length to its reference.

\section{Application to paleoclimate data sets} \label{sec:pcts}

We will now apply the different phase space reconstruction methods in combination with recurrence network
analysis to two well-studied paleoclimate time series from Ecuador and Mexico. The aim is to detect nonlinear regime 
shifts in past climate variability. To do so, we will first introduce the proxy time series and
the exact analysis procedure before interpreting the results and drawing 
conclusions about the suitability of methods on the one hand and on the variability of the regional climate
of the past $2000$ years on the other hand.

\subsection{Data}

\begin{figure}
\centering
\includegraphics[width=0.99\linewidth]{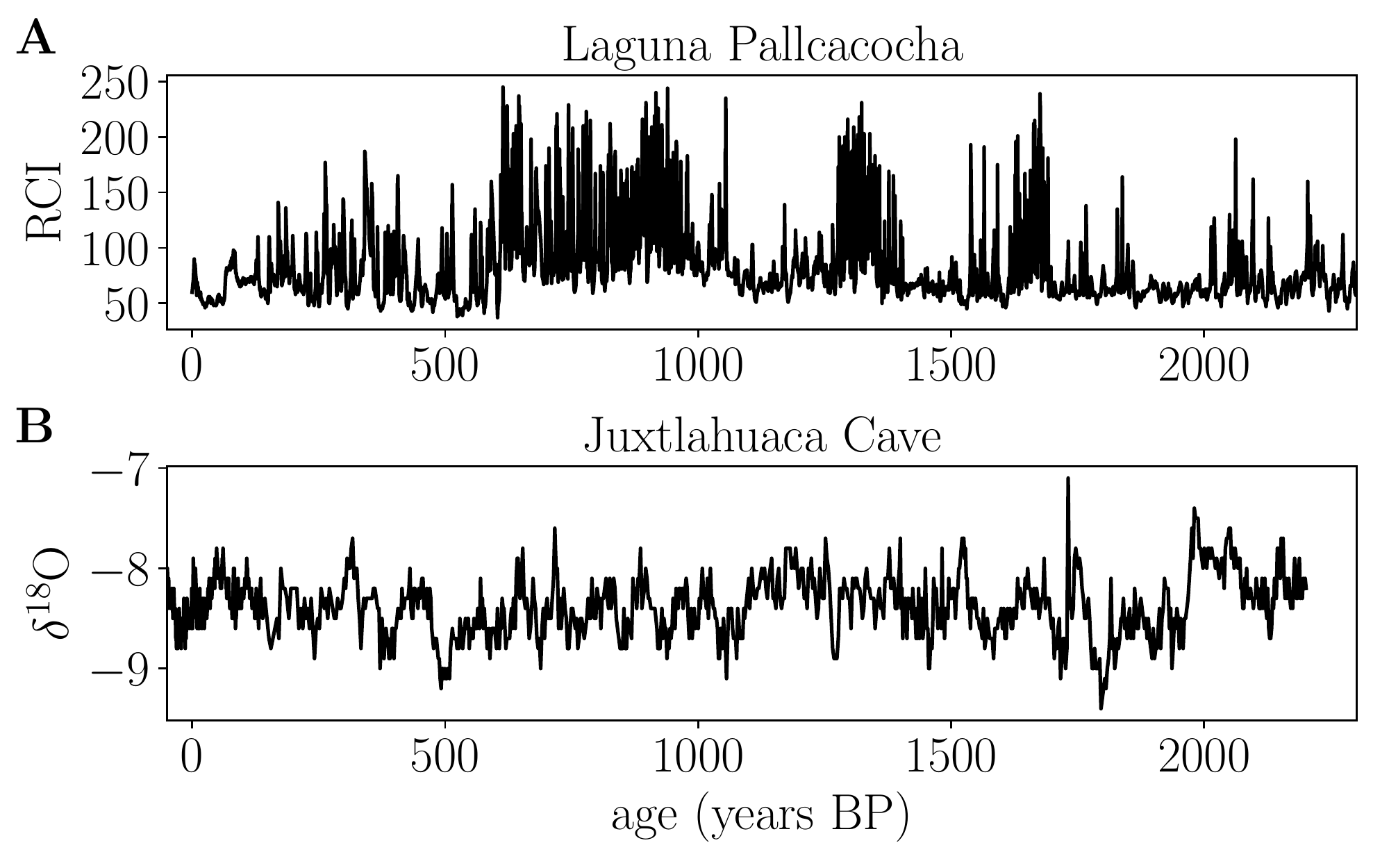}
\caption{Paleoclimate data from (A) Laguna Pallcacocha~\cite{Moy2002} (red color intensity) and (B) Juxtlahuaca cave~\cite{Lachniet2012} (oxygen-18 residuals in units of \permil PDB). Data provided by https://www.ncdc.noaa.gov/.}
\label{fig:paleo-ts}
\end{figure}
The first data set (Fig.~\ref{fig:paleo-ts}A) comprises the most recent $2300$ years ($N = 4683$) of the 
red color intensity (RCI) data from Laguna Pallcacocha, a high-altitude lake sediment from southern Ecuador. 
The red color intensity serves as a proxy for rainfall intensity and has been related to the strength
of historical El Ni\~{n}o Southern Oscillation (ENSO) events~\cite{Moy2002}. 
The data have a high temporal resolution with an average sampling time of 
$\left\langle\Delta t\right\rangle = 0.49$ 
years with  standard deviation
$\sigma = 0.27$ years. The values of the red color intensity vary between $37$ and $245$ with mean 
$\mu_{\text{RCI}} = 96.7$ and standard deviation $\sigma_{\text{RCI}} = 42.9$. High values of the RCI 
correspond to light colored laminae in the sediment and are associated with strong rainfall events 
triggered by moderate to strong warm ENSO episodes (El Ni\~{n}o events)~\cite{Moy2002}.

The second data set (Fig.~\ref{fig:paleo-ts}B) consists of oxygen isotope ratios ($\delta^{18}$O) of the 
past $2250$ years ($N = 1218$) recorded in a stalagmite from Juxtlahuaca cave
in the Basin of Mexico~\cite{Lachniet2012}. This data set has also been related to rainfall variability in 
that region mostly affected by the strength of the North American summer monsoon. The latter is known to be
affected by ENSO, where warm ENSO episodes lead to a weakened summer monsoon in this area and vice versa. 
This second data set has a larger average sampling time of $\left\langle\Delta t\right
\rangle = 1.8$\,years with standard deviation $\sigma = 0.86$\,years.
The proxy values vary between $-9.4$\,\permil PDB (Peedee Belemnite) and $-7.1$\,\permil PDB with mean 
$\mu_{\delta^{18} \text{O}} = -8.4$\,\permil PDB and standard 
deviation $\sigma_{\delta^{18} \text{O}} = 0.29$\,\permil PDB. More negative values indicate higher rainfall 
amounts and less negative values less rainfall in the study area~\cite{Lachniet2012}.

\subsection{Analysis procedure}

We analyze both data sets using windowed recurrence network analysis. That is, after reconstructing the 
phase space of the system, we take the first $W$ points of the time series, construct a recurrence 
network as described in Sec.~\ref{subsec:rna} and calculate the transitivity of the network. 
We focus here on transitivity as the only network measure, since the results shown in Sec.~\ref{sec:ms}
for the two model systems indicate that the obtained results could be slightly more robust than for the average
path length. Then we move
the window and repeat the analysis, such that we end up with a time series of network measures. We assign
the most recent time of the $W$ data points to the network measure as it is the result of the system
dynamics \emph{up to} this point. In order to
characterize at which points in time the system exhibits extraordinary behavior, we use
surrogate data to estimate significance levels~\cite{Schreiber2000}. Our surrogates are created by randomly 
drawing $W$ embedded state vectors from the time series and calculating the transitivity of the 
corresponding recurrence network. This
is repeated $N_{\text{surr}} = 500$ times and two-sided $95\%$ significance levels are estimated using 
the $2.5$th and the $97.5$th percentile of the resulting network measures for the surrogates.
This kind of analysis has already been shown to deliver valuable insights into the variability of the 
climate of the past~\cite{Donges2011,Donges2011a,Donges2015a}. 
It should be noted that the analysis of recurrence networks 
itself does not require regular sampling but can handle non-uniformly sampled data. 
Still, the reconstruction of the phase space is required prior 
to this analysis. As we have seen in the previous section that the results from interpolated time series
perform quite well and because the interpretation of the results is not very clear for 
non-uniformly sampled data, we here use interpolated data.

We perform the windowed recurrence network analysis for different window 
widths, embedding dimensions and different methods to reconstruct the phase space. For each of the methods,
we vary the internal embedding parameter, i.\,e. the delay time $\tau$ for time delay embedding and the 
number of neighboring points included when estimating the derivatives using discrete Legendre polynomials 
and MoTaBaR to check the robustness of the results. We chose to use linear interpolation for time delay 
embedding to get the required regular sampling of the data and compare the results to those obtained with 
differential embedding. According to the results obtained for the phase space reconstruction for the model 
systems (Sec.~\ref{sec:ms}), we use spline interpolation and scale the data to unit 
variance before constructing the recurrence networks for discrete Legendre polynomials.
For MoTaBaR, we use the internal interpolation routine and also scale the data to unit variance. We vary the 
embedding dimension between $3$ and $6$ and choose window lengths of $W = \left[100,150,200,250,500\right]$. 
For time delay embedding, we vary the delays between $0$ and $50$ for Laguna Pallcacocha and between
$0$ and $100$ for Juxtlahuaca cave. For differential embedding, we
consider values of $p$ between $2$ and $20$ and for MoTaBaR, we vary $N_{mtb}$ between $10$ and $50$. 

\subsection{Results}

\begin{figure*}
\centering
\includegraphics[width=0.86\linewidth]{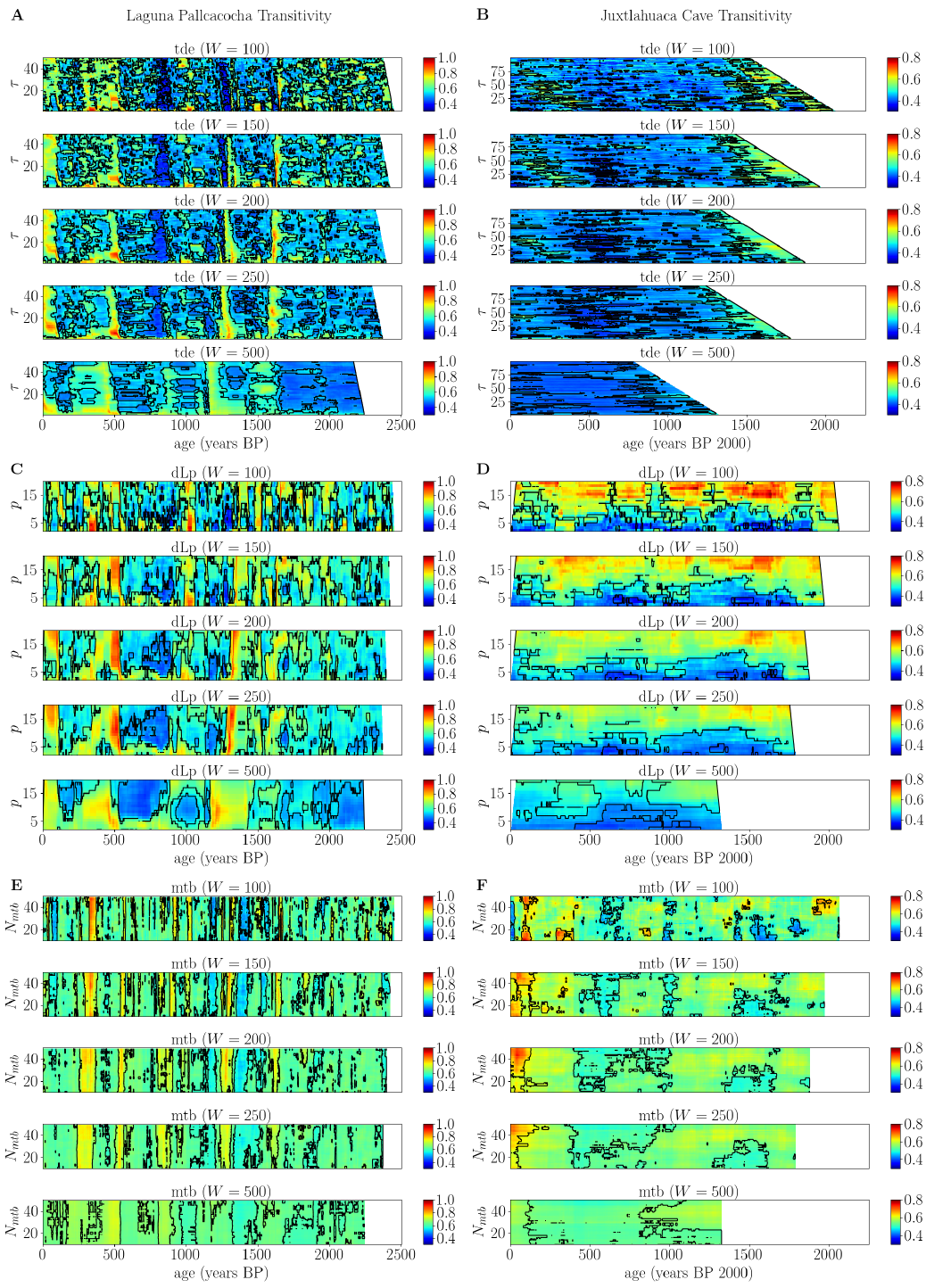}
\caption{Network transitivity $\mathcal{T}$ as a function of age of Laguna Pallcacocha (left panels) and Juxtlahuaca cave (right panels) data sets for embedding dimension $m=4$, different phase space reconstructions and different window widths $W$. (A,B) Time delay embedding for varying delay times $\tau$. (C,D) Differential embedding with discrete Legendre polynomials and different values of the number of neighboring points $p$. (E,F) Differential embedding using MoTaBaR when varying the number of included points $N_{mtb}$.}
\label{fig:paleo}
\end{figure*}

Figure~\ref{fig:paleo} shows the results for the two data sets and three methods mentioned above 
for embedding dimension $m = 4$.
This dimension has been chosen in accordance with the false nearest neighbor criterion~\cite{Kennel1992}.
In fact, when further increasing the embedding dimension, we do not observe significant changes in
the results. 

For Laguna Pallcacocha and time delay embedding (Fig.~\ref{fig:paleo}A), we identify four to five periods in 
which the network transitivity is significantly above that of the surrogate data. 
Also, we identify one to two periods with significantly lower transitivity. We observe that the timing of 
those anomalies shifts to more recent times when increasing the window width. This can be understood when 
recalling that we assign the most recent time to our time windows.
For differential embedding using discrete Legendre polynomials to estimate the derivatives 
(Fig.~\ref{fig:paleo}C), we find similar but not identical significant periods. 
In general, the significant periods are wider, in particular for higher values of $W$ such 
that there is no clear separation between some periods like for time delay embedding. 
Also, the percentage of windows with 
significant values of the transitivity is higher than for time delay embedding. 
For differential embedding using MoTaBaR (Fig.~\ref{fig:paleo}E), we observe a 
rather stable behavior when varying the number of utilized points,
but we also observe more and shorter significant regimes with different timings than those identified
by the other two methods. Additionally, the minimum/maximum values of the network transitivity are 
larger/smaller than those obtained with time delay embedding and differential embedding using discrete 
Legendre polynomials. 

\begin{figure}
\centering
\includegraphics[width=0.99\linewidth]{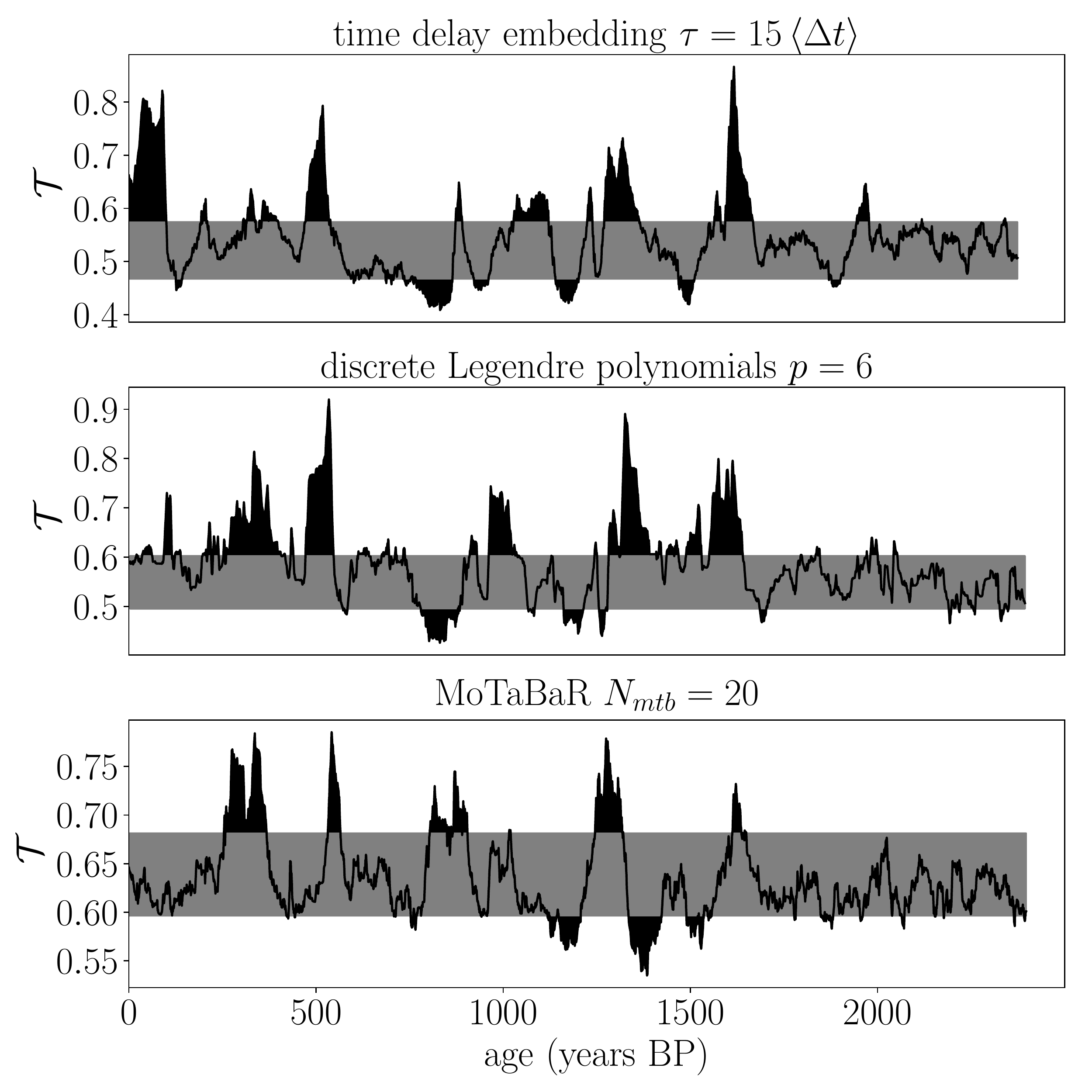}
\caption{Network transitivity $\mathcal{T}$ as a function of age for the Laguna Pallcacocha data set using embedding dimension $m=4$, window width $W=200$ and different phase space reconstruction methods. For time delay embedding, the delay time is $\tau=15 \left\langle \Delta t\right\rangle$; for differential embedding with discrete Legendre polynomials, the number of neighboring points is $p=6$; and for differential embedding using MoTaBaR, the number of included points is $N_{mtb}= 20$.}
\label{fig:paleo-single}
\end{figure}

The aforementioned behavior and the different timing of the identified significant periods is highlighted 
in more detail in Fig.~\ref{fig:paleo-single}. 
Here, we choose the delay time to be $\tau=15 \left\langle \Delta t\right\rangle$. This choice is in 
accordance with the first zero of the autocorrelation function of the time series using an estimator that can handle 
non-uniformly sampled data~\cite{Rehfeld2011}. 
For the discrete Legendre polynomials, we choose $p=6$. The estimator from
Gibson suggests $p=2$, but in order to also estimate the higher order derivatives, we prefer to choose a 
slightly larger number of neighbors as already discussed in Sec.~\ref{sec:ms}. 
For MoTaBaR, the results do not vary a lot depending on the number
of included points, so we decided to have a closer look at $N_{mtb}= 20$.
We see that the results for time delay embedding and discrete Legendre polynomials generally are in good 
agreement though some periods are more pronounced in either of the methods. 
For MoTaBaR, we partially find significant deviations from the surrogates at the same time as for the other methods, 
but we also observe shifts in timing and even opposite behavior as for example around $900$ years before
present (BP). This highlights that it is not enough to 
consider a single analysis method or a single analysis parameter to infer information about the
variability of the past climate. Instead, to gain confidence in the analysis results, several methods 
and a broad range of analysis parameters should be considered and checked for robustness.

The Laguna Pallcacocha RCI data set has previously been related to the succession of moderate to strong
El Ni\~{n}o events. Moy et al.~\cite{Moy2002} found that the frequency of those events has increased after
$7000$\,years BP and declined again from about $1200$\,years BP onward. 
Here we do not analyze the number of individual El Ni\~{n}o events 
but rather aim at detecting general nonlinear changes and dynamical regime shifts in the climate signal
represented by the considered proxy. Periods such
as the Little Ice Age (LIA) or the Medieval Climate Anomaly (MCA) have been well expressed in North Atlantic climate
proxies and Northern Hemisphere temperature trends have been argued to also influence ENSO and the
position of the Intertropical Convergence zone (ITCZ) and, thus,
Southern Hemisphere climate~\cite{Novello2016}. Starting from the past, we find consistent anomalies
in recurrence network transitivity estimated using
time delay embedding and discrete Legendre polynomials in the periods from (i) $2000-1950$\,years 
BP, (ii) $1650-1550$\,years BP, (iii) $1380-1275$\,years BP, (iv) $870-750$\,years BP, (v) $540-470$\,years 
BP, (vi) $400-300$\,years BP, and (vii) $100-0$\,years BP. Generally, the anomalies start and finish 
slightly earlier for differential embedding using discrete Legendre polynomials than for time delay
embedding. The first anomalies (i) and (ii) are similarly pronounced for both methods and
might be related to the Roman Warm Period in Europe. This episode of rather stable warm temperatures 
might also have affected Southern Hemisphere climate by influencing SST temperatures in the Pacific 
leading to changes in ENSO. The latter is also visible in the MoTaBaR 
results. Anomaly (iii) is rather pronounced and might relate to changes in ENSO frequency around 
$1200$\,years BP as observed by Moy et al~\cite{Moy2002}. Anomaly (iv) shows a significantly reduced 
network transitivity and co-occurred with the termination of the MCA. 
Vuille et al. determined this episode to be around $1050-850$\,years BP in the Southern 
Hemisphere~\cite{Vuille2012}. The MCA in South America is associated with
dryer climate due to a weakened South American monsoon system responding to anomalously high
Northern Hemisphere temperatures.
During this interval, we cannot observe consistent significant signals in the 
network transitivity across the different methods. Still, the anomaly we observe between 
$870-750$\,years BP is likely the result of a change towards more complex climate variability
as compared to the relatively stable climate during the MCA. 
Anomalies (v) and (vi) show significantly enhanced network 
transitivity and can be associated with the LIA that has been found to be around 
$350-130$\,years BP in that area~\cite{Vuille2012}. Finally, anomaly (vii) can be
related to the onset of the Current Warm Period and the associated recent climate change.
Also, it should be mentioned that we find two periods where we observe inconsistencies between time delay
embedding and differential embedding using discrete Legendre polynomials, one around $1500$\,years
BP and the other one around $1000$\,years BP. For those periods, we cannot draw any conclusions about
whether something significant happened, even though for the second time interval, we might expect to 
find signatures of the onset of the MCA. 

Unlike for the Laguna Pallcacocha record, for the Juxtlahuaca cave, we generally cannot identify
consistent periods during which the network transitivity shows 
significant anomalies over the full considered parameter range. For time delay embedding 
(Fig.~\ref{fig:paleo}B), we observe some elevated values in the very past and during more recent times 
together with some significantly reduced values around $500$ to $700$ years BP, 
but the timing varies a lot with both, the delay time $\tau$ and window width $W$. 
For differential embedding using discrete Legendre polynomials (Fig.~\ref{fig:paleo}D), we find that for 
higher values of $p$, almost all values of the transitivity are significantly above those of the surrogate 
data. This may be understood when recalling that higher values of $p$ increasingly smoothen the data
corresponding to a low-pass filter. At some value of $p$ that also depends on the sampling rate, the
time series becomes too flat, which is probably what happens here. For lower values of $p$ we observe a similar
pattern as for time delay embedding, which is for some window widths also more localized. For differential 
embedding using MoTaBaR (Fig.~\ref{fig:paleo}F), we also find some significantly reduced values of the 
transitivity at similar times as for the other methods.
Although these seem to be more robust than the other results, they appear still much 
less reliable than the results obtained for the Laguna Pallcacocha data set. 
Also, we observe again, that the values of the transitivity obtained with MoTaBaR are generally higher than 
those obtained with the other methods but in this case show some very high but no very low values. 

The Juxtlahuaca cave record has been previously used to reconstruct rainfall over the past $2500$\,years 
in the Basin of Mexico
and related drought conditions to cultural change in this area. It has been argued that
not only changes in mean conditions but also changes in climate variability can 
foster cultural change~\cite{Donges2011a,Donges2015a}. However, from the perspective of recurrence analyses, 
we cannot make any reliable statements for this data set.
In turn, other more suitable records need to be obtained and analyzed to further examine the
aforementioned relation.

\section{Conclusions} \label{sec:conclusion}

The reconstruction of a system's phase space from measured data is an important step in the nonlinear 
analysis of real-world time series in many fields of application. However, care has to be taken
in choosing an appropriate embedding, in particular when the data are non-uniformly sampled and noisy. 
For two paradigmatic model systems, we have used recurrence network measures to systematically compare the 
reference attractor to system attractors reconstructed using time delay embedding and differential 
embedding for different methods to estimate derivatives. When varying the noise level of the data 
and the shape of the distribution of the non-uniform sampling intervals, we found that 
differential embedding using discrete Legendre polynomials or Moving Taylor Bayesian Regression can be 
an alternative to time delay embedding for reconstructing the phase space of a system. 

We also studied two precipitation sensitive paleoclimate proxy records from Ecuador and Mexico by
combining windowed recurrence network analysis and different methods for phase space reconstruction.  
We found that the Laguna Pallcacocha data set from southern Ecuador gives robust results when varying
embedding and network analysis parameters and therefore seems to be well-suited for such an 
analysis. Anomalies detected are in general in good agreement for two of the three considered 
methods but partially differ significantly for the third, highlighting that different
phase space reconstructions should be compared before interpreting the results. 
We could relate the detected anomalies to known episodes of past climate change such as the Medieval 
Climate Anomaly and the Little Ice Age. Our corresponding results are consistent with previous work
demonstrating that those periods, even though most pronounced in the Northern Hemisphere, have also left 
significant imprints in past climate variability of the Southern Hemisphere.
For the Juxtlahuaca cave data set
from southern Mexico, we could not identify robust anomalies using any of the methods applied.
Thus, we have to conclude that this data set does not 
seem to be suitable to make any well-grounded statements about the variability of rainfall and possible 
regime shifts when using windowed recurrence network analysis.
In turn, this shows that not all paleoclimate data sets are suited for this kind of analysis,
and care has to be taken when analyzing paleoclimate time series with recurrence networks and related
techniques.
It is necessary to test the results for robustness by varying the different analysis parameters and 
comparing the results to draw reliable conclusions.

When respecting these insights for analyzing non-uniformly and noisy data in the paleoclimatologic 
context, we find that a framework of different phase space reconstruction methods combined with windowed 
recurrence network analysis and the associated robustness tests can be used to gain valuable insights into
past climate variability. Thus, it should be further used to better understand and
systematically study the climate of the past as this might help to better anticipate possible future 
changes. 

\begin{acknowledgments}
This work has been financially supported by the German Federal Ministry for Education and Research (BMBF)  
via the BMBF Young Investigators Group \emph{CoSy-CC$^2$ -- Complex Systems Approaches to Understanding  
Causes and Consequences of Past, Present and Future Climate Change} (grant no. 01LN1306A) and the
Studienstiftung des deutschen Volkes. Calculations have been performed with the help of the Python packages 
\texttt{pyunicorn}~\cite{Donges2015} and \texttt{MoTaBaR}~\cite{Heitzig2013}.
\end{acknowledgments}

\appendix*
\section{Estimating derivatives using discrete Legendre polynomials} \label{app-dLp}

In order to derive the formula for estimating the derivatives of a time series using discrete Legendre polynomials (Eq.~\eqref{eq:dLp}), we assume 
a discrete set of one-dimensional data $\left\lbrace x_i\right\rbrace_{i=1}^N$ with irregular 
sampling intervals $\Delta t_{i,n} = t_{i+n} - t_i$. We can apply a 
discrete linear filter
\begin{align}
\omega_j (t_i) &= \sum_{n=-p}^p r_{j,p,n}^{(i)} x(t_i+\Delta t_{i,n}) \label{eq:filtered}
\end{align}
to the data and perform a Taylor expansion for small $\Delta t_{i,n}$ yielding
\begin{align}
\omega_j (t_i) &= \sum_{n=-p}^p r_{j,p,n}^{(i)} \left[\sum_{\nu =0}^{\infty} \dfrac{(\Delta t_{i,n})^{\nu}}{\nu !} \dfrac{d^{\nu}}{d t^{\nu}} x_i\right]\nonumber\\
&= \sum_{\nu =0}^{\infty} \dfrac{1}{\nu !} \dfrac{d^{\nu}}{d t^{\nu}} x_i \sum_{n=-p}^p (\Delta t_{i,n})^{\nu} r_{j,p,n}^{(i)}.  \label{eq:taylor-expanded}
\end{align}
As we aim at deriving an expression for $\omega_j (t_i)$ that is proportional to the $j$th derivative,
we have to choose the filter function $r_{j,p,n}^{(i)}$ such that it is orthogonal to 
$(\Delta t_{i,n})^{\nu}$ for $\nu < j$:
\begin{align}
\sum_{n=-p}^p (\Delta t_{i,n})^{\nu} r_{j,p,n}^{(i)}  &= 0. \label{eq:orth-cond}
\end{align}
It can be shown that the discrete Legendre polynomials defined by the recursion~\eqref{eq:rec-relation}
are mutually orthonormal filters, that is, they satisfy
\begin{align}
\sum_{n=-p}^p r_{k,p,n}^{(i)} r_{l,p,n}^{(i)} &= \delta_{k,l} \nonumber
\end{align}
with $\delta_{k,l}$ being the Kronecker Delta. 
Thus, condition~\eqref{eq:orth-cond} holds and the Taylor expansion~\eqref{eq:taylor-expanded} reduces to 
\begin{align}
\omega_j (t_i) &= \dfrac{1}{j !} \dfrac{d^{j}}{d t^{j}} x_i \sum_{n=-p}^p (\Delta t_{i,n})^j r_{j,p,n}^{(i)} \nonumber
\end{align}
to leading order. Equating this result with Eq.~\eqref{eq:filtered} yields
the desired result~\eqref{eq:dLp}
\begin{align}
\dfrac{d^j}{d t^j} x_i &\approx \dfrac{j !}{c_{j,p}(\Delta t_{i,n})} \sum_{n=-p}^p r_{j,p,n}^{(i)} x_{i+n}, \nonumber
\end{align}
relating the $j$th derivative of $x_i$ to the data points 
$\left\lbrace x_{i+\mu}\right\rbrace_{\mu = -p}^p$ of the time series. For given $p$, the discrete 
Legendre polynomials $r_{j,p,n}^{(i)}$ can easily be calculated numerically using the recursive
relation~\eqref{eq:rec-relation} and the normalization condition~\eqref{eq:normalization}.

\end{document}